\documentclass{article}
\usepackage{breckenridge}
\usepackage{epsfig}
\frompage{000} \topage{000}                                              
\def\rightmark{Hadrons in Nuclei}
\def\leftmark{U. Mosel}

\title{Hadrons in Nuclei}
\authors{
{Ulrich Mosel$^a$
{}} \\[2.812mm]
{\normalsize
Institut fuer Theoretische Physik, Universitaet Giessen\\
D-35392 Giessen, Germany\\[0.2ex]
}}

\abstract{Changes of hadronic properties in dense nuclear
matter as predicted by theory have usually been investigated by
means of relativistic heavy-ion reactions. In this talk I show
that observable consequences of such changes can also be seen in
more elementary reactions on nuclei. Particular emphasis is put on
a discussion of photonuclear reactions; examples are the dilepton
production at $\approx 1$ GeV and the hadron production in nuclei
at 10 - 20 GeV photon energies. The observable effects are
expected to be as large as in relativistic heavy-ion collisions
and can be more directly related to the underlying hadronic
changes. }


\begin{document}

\maketitle
\setcounter{page}{1}

\section{Introduction}\label{intro}

That hadrons can change their properties and couplings in the
nuclear medium has been well known to nuclear physicists since the
days of the Delta-hole model that dealt with the changes of the
properties of the pion and Delta-resonance inside nuclei
\cite{Ericsson-Weise}. Due to the predominant $p$-wave interaction
of pions with nucleons one observes here a lowering of the pion
branch with increasing pion-momentum and nucleon-density. A direct
observation of this effect is difficult because of the strong
final state interactions (in particular absorption) of the pions.

More recently, in-medium changes of vector mesons have found
increased interest, mainly because these mesons couple strongly to
the photon so that electromagnetic signals could yield information
about properties of hadrons deeply embedded into nuclear matter.
Indeed, the CERES experiment \cite{CERES} has found a considerable
excess of dileptons in an invariant mass range from $\approx 300$
MeV to $\approx 700$ MeV as compared to expectations based on the
assumption of freely radiating mesons. This result has found an
explanation in terms of a shift of the $\rho$ meson spectral
function down to lower masses, as expected from theory (see, e.g.,
\cite{Peters} \cite{Post}).

However, the actual reason for the observed dilepton excess is far
from clear. Both models that just shift the pole mass of the
vector meson as well as those that also modify the spectral shape
have successfully explained the data; in addition, even a
calculation that just used the free radiation rates with their --
often quite large -- experimental uncertainties was compatible
with the observations \cite{Koch}. There are also calculations
that attribute the observed effect to radiation from a quark-gluon
plasma \cite{Renk}.

I have therefore already some years ago proposed to look for the
theoretically predicted changes of vector meson properties inside
the nuclear medium in reaction on normal nuclei with more
microscopic probes \cite{Hirschegg}. Of course, the nuclear
density felt by the vector mesons in such experiments lies much
below the equilibrium density of nuclear matter, $\rho_0$, so that
naively any density-dependent effects are expected to be smaller
than in heavy-ion reactions.

On the other hand, there is advantage to these experiments: they
proceed with the spectator matter being close to its equilibrium
state. This is essential because all theoretical predictions of
in-medium properties of hadrons are based on an equilibrium model
in which the hadron (vector meson) under investigation is embedded
in cold nuclear matter in equilibrium and with infinite extension.
However, a relativistic heavy-ion reaction proceeds -- at least
initially -- far from equilibrium. Even if equilibrium is reached
this state changes by cooling through expansion and particle
emission and any observed signal is built up by integrating over
the emissions from all these different stages of the reaction.

In this paper I summarize results that we have obtained in studies
of observable consequences of in-medium changes of hadronic
spectral functions in reactions of elementary probes with nuclei.
I demonstrate that the expected in-medium sensitivity in such
reactions is as high as that in relativistic heavy-ion
collisions and that in particular photonuclear reactions present
an independent, cleaner testing ground for assumptions made in
analyzing heavy-ion reactions.

\section{Dilepton Production}

\begin{figure}[h]
\centering{\epsfig{figure=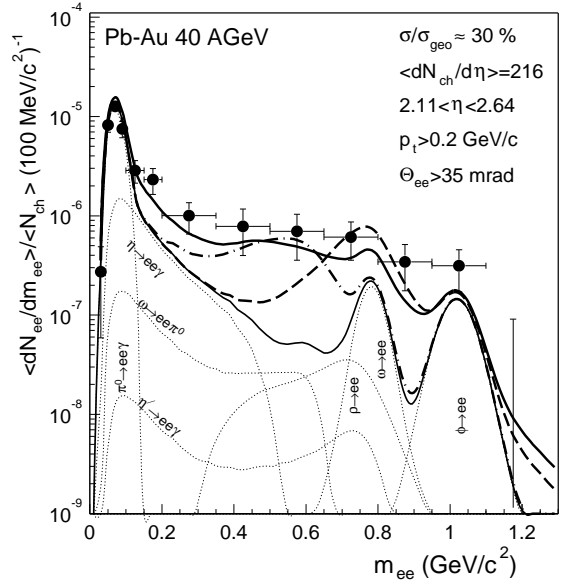,width=8cm}}
 \vspace*{-0.2cm} \caption[]{Invariant  dilepton mass spectrum
 obtained with the CERES experiment in Pb + Au collisions at 40
 AGeV (from \cite{CERES}). The thin curves give the contributions
 of individual hadronic sources to the total dilepton yield, the
 fat solid (modified spectral function) and the dash-dotted
 (dropping mass only) curves give the results of calculations
 employing an in-medium modified spectral function of the vector
 mesons.} \label{CERES}
\end{figure}

The CERES experiment has received a lot of publicity for its
observation of an excess of dileptons with invariant masses below
those of the lightest vector mesons \cite{CERES}. Explanations of
this excess have focussed on a change of in-medium properties of
these vector mesons in dense nuclear matter. The radiating sources
can be nicely seen in Fig.~\ref{CERES} that shows the latest
dilepton spectrum obtained in a rather low-energy run at 40 AGeV.

The figure exhibits clearly the rather strong contributions of the
vector mesons -- both direct and through their Dalitz decay -- at
invariant masses above about 500 MeV. If this strength is shifted
downward, caused by an in-medium change of the vector-meson
spectral functions, then the observed excess can be explained as
has been shown by various authors (see e.g.\ \cite{Brat-Cass}).

As mentioned above such explanations always suffer from an
inherent inconsistency: while the observed signal integrates over
many different stages of the collision -- nonequilibrium and
equilibrium, the latter at various densities and temperatures --
the theoretical input is always calculated under the assumption of
a vector meson in nuclear matter in equilibrium. We have therefore
looked for possible effects in reactions that proceed much closer
to equilibrium and have thus studied the dilepton production in
photon-induced reactions on the nucleus. It is not \emph{a priori}
hopeless to look for in-medium effects in ordinary nuclei: Even in
relativistic heavy-ion reactions that reach baryonic densities of
the order of 3 - 10 $\rho_0$ many observed dileptons actually stem
from densities that are much lower than these high peak densities.
Transport simulations have shown \cite{Brat-Cass} that even at the
CERES energies about 1/2 of all dileptons come from densities
lower than $2 \rho_0$. This is so because in such reactions the
pion-density gets quite large in particular in the late stages of
the collision, where the baryonic matter expands and its density
becomes low again. Correpondingly many vector mesons are formed
late in the collision (through $\pi + \pi -> \rho$) and their
decay to dileptons thus happens at low baryon densities.

The calculations are done in a combination of coherent initial
state interactions that lead to shadowing at photon energies above
about 1 GeV and incoherent final state interactions. The shadowed
incoming photon produces, for example, a vector meson which then
cascades through the nucleus. The latter process we describe by
means of a coupled-channel transport theory. The details are
discussed in ref. \cite{Effe}. A new feature of these calculations
is that vector mesons with their correct spectral functions can
actually be produced and transported consistently. This is quite
an advantage over earlier treatments in which the mesons were
always produced and transported with their pole mass and their
spectral function was later on folded in only for their decay.

\begin{figure}[h]
\vspace*{-0.5cm}
\centering{\epsfig{figure=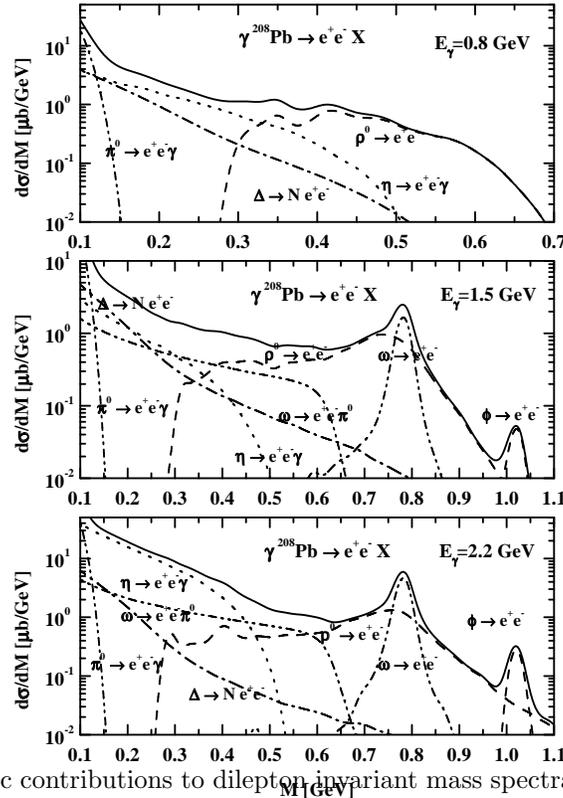,width=9cm}} \vspace*{-2cm}

\caption{Hadronic contributions to dilepton
invariant mass spectra for $\gamma + ^{208}Pb$ at the three photon
energies given (from \cite{Effe}). Compare with Fig.~\ref{CERES}.}
\label{Fige+e-}
\end{figure}

\begin{figure}[htb]
\centering{\epsfig{figure=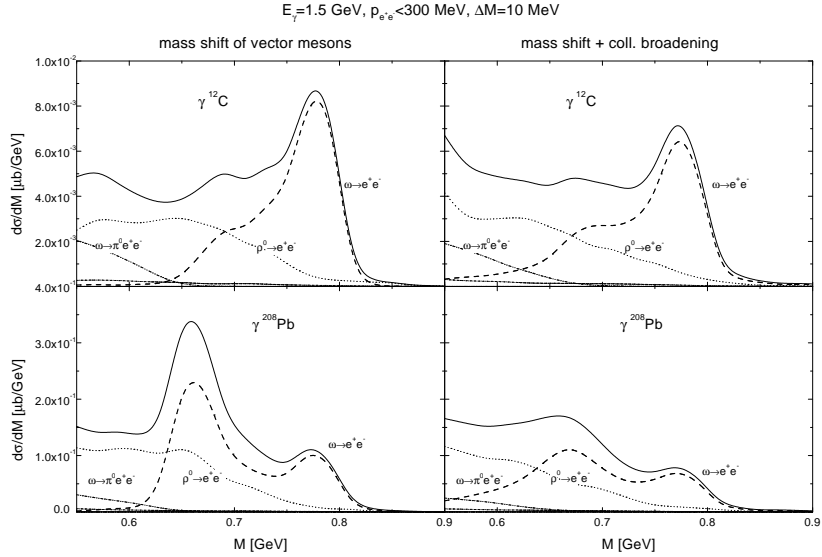,width=11.5cm}}
\vspace*{-0.5cm} \caption{Dilepton mass yield with a
dilepton-momentum cut of 300 MeV. Shown on the left are results of
a calculation that uses only a shift of the pole mass of the
vector mesons. On the right, results are given for a calculation
using both mass shift and collisional broadening (from
\cite{Effe}).} \label{Figdlim}
\end{figure}

A typical result of such a calculation for the dilepton yield --
after removing the Bethe-Heitler component -- is given in Fig.
\ref{Fige+e-}.
Comparing this figure with Fig. \ref{CERES} shows
that exactly the same sources, and none less, contribute to the
dilepton yield in a photon-induced reaction at 1 - 2 GeV photon
energy as in relativistic heavy-ion collisions at 40 AGeV!
The question now remains if we can expect any observable effect of
possible in-medium changes of the vector meson spectral functions
in medium in such an experiment on the nucleus where -- due to
surface effects -- the average nucleon density is below $\rho_0$.
This question is answered by the results of Fig.\ \ref{Figdlim}.
This figure shows the dilepton spectra to be expected if a
suitable cut on the dilepton momenta is imposed. In the realistic
case shown on the right, which contains both a collision
broadening and a mass shift, it is obvious that a major signal is
to be expected: in the heavy nucleus $Pb$ the $\omega$-peak has
completely disappeared from the spectrum. The sensitivity of such
reactions is thus as large as that observed in ultrarelativistic
heavy-ion reactions.

An experimental verification of this prediction would be a major
step forward in our understanding of in-medium changes\footnote{An
experiment at JLAB is under way \cite{Weygand}.}. It would
obviously present a purely hadronic base-line to all data on top
of which all 'fancier' explanations of the CERES effect in terms
of radiation from a QGP and the such would have to live.

\section{Hadron Formation}
A major experimental effort at RHIC experiments has gone into the
observation of jets in ultrarelativistic heavy-ion collisions and
the determination of their interaction with the surrounding quark
or hadronic matter \cite{Jet}. Such experiments are obviously very
sensitive to hadron formation times. In addition they can yield
information on interactions while the final hadron is still being
formed.

A complementary process is given by the latest HERMES results at
HERA for photon-induced hadron production at high energies
\cite{Hermes}. Here the photon-energies are of the order of 10 -
20 GeV, with rather moderate $Q^2 \approx 2$ GeV. Again, the
advantage of such experiments is that the nuclear matter with
which the interactions happen is at rest and in equilibrium.

In the high-energy regime the shadowing of the incoming photon,
which is due to interference between interactions of the incoming
bare photon and its hadronic components with the nucleons, becomes
important. This coherence in the incoming state has to be combined
with the incoherent treatment of the final state interactions in
transport theory. For this purpose T.\ Falter has derived a novel
expression for incoherent particle production on the nucleus
\cite{falterinc} that allows for a clean-cut separation of the
coherent initial state and the incoherent final state interactions
which we again treat with our coupled-channel transport theory.
\begin{figure}[h]
\vspace*{-0.0cm}
 \centering{\epsfig{figure=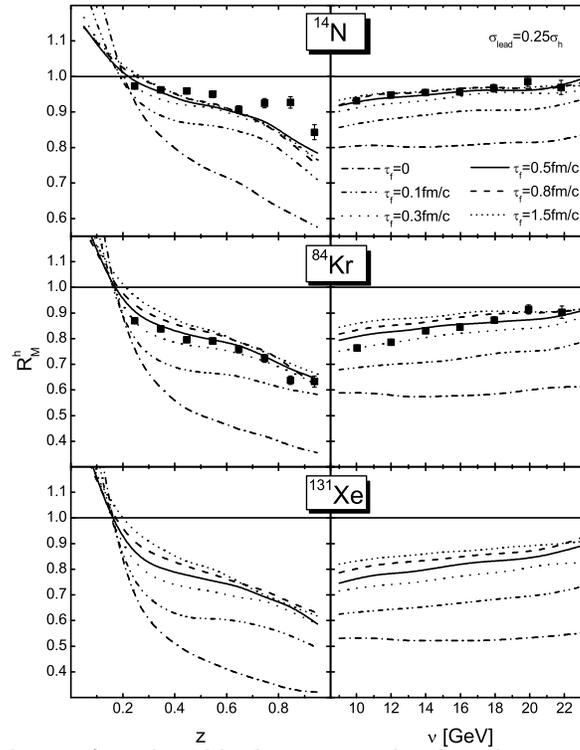,width=8.0cm}}
\vspace*{-0.5cm}

\caption{Multiplicity of produced hadrons normalized to the proton
as a function of photon-energy $\nu$ (right) and of the energy of
the produced hadrons relative to the photon-energy, $ z =
E_h/\nu$. The curves are calculated for different formation times
given in the figure (from \cite{falterform}).} \label{formtime}
\end{figure}

An example of the results obtained is given in Fig.\
\ref{formtime}. The figure clearly shows that the observed hadron
multiplicities can be described only with formation times $>
\approx 0.3$ fm. The curves obtained with larger formation times
all lie very close together. This is a consequence of the finite
size of the target nucleus: if the formation time is larger than
the time needed for the preformed hadron to transverse the nucleus
then the sensitivity to the formation time is lost. In
\cite{falterform} we have also shown that the $z$-dependence on
the left side exhibits some sensitivity to the interactions of the
leading hadrons during the formation; the curves show in Fig.\
\ref{formtime} are obtained with a leading hadron cross section of
$0.25 \sigma_h$, where $\sigma_h$ is the 'normal' hadronic
interaction cross section.

\section{Conclusions}\label{concl}
In this talk I have illustrated with the help of two examples that
photonuclear reactions can yield information that is important and
relevant for an understanding of high density--high temperature
phenomena in ultrarelativistic heavy-ion collisions. I have shown
that the expected sensitivity of dilepton spectra in photonuclear
reactions in the 1 - 2 GeV range is as large as that in
ultrarelativistic heavy-ion collisions. I have also illustrated
that the analysis of hadron production spectra in high-energy
electroproduction experiments at HERMES gives information about
the interaction of forming hadrons with the surrounding hadronic
matter. This is important for any analysis that tries to obtain
signals for a QGP by analysing high-energy jet formation in
ultrarelativistic heavy-ion reactions.

\section*{Acknowledgement}
This talk is based on work done together with Martin Effenberger,
Thomas Falter and Kai Gallmeister. The work on which it is based
has been supported by the Deutsche Forschungsgemeinschaft, the
BMBF and GSI Darmstadt.

\begin{notes}
\item[a] E-mail: mosel@physik.uni-giessen.de
\end{notes}

\vfill\eject
\end{document}